\pdfoutput=1
\documentclass[prc,preprintnumbers,eqsecnum,floats,aps,showpacs,twocolumn,floatfix,superscriptaddress]{revtex4}
\usepackage{graphicx}

\newcommand{\mc}[1]{\multicolumn{#1}}
\newcommand{\tautau}{\vec{\tau_1} \vec{\tau_2}}
\newcommand{\sst}[1]{\scriptscriptstyle{#1}}
\newcommand{\ignore}[1]{}
\def \im {{\rm i}}
\def \vcks {{ g_{\sst \Lambda {\rm N} K^*}^{\sst \rm V}}}
\def \tcks {{g_{\sst \Lambda {\rm N} K^*}^{\sst \rm T}}}
\def \vco {{g_{\sst {\rm N N} \omega}^{\sst \rm V}}}
\def \tco {{g_{\sst {\rm N N} \omega}^{\sst \rm T}}}
\def \vcr {{g_{\sst {\rm N N} \rho}^{\sst \rm V}}}
\def \tcr {{g_{\sst {\rm N N} \rho}^{\sst \rm T}}}
\def \ce {{g_{\sst {\rm N N} \eta}}}
\def \alfao {\alpha_{\omega}}
\def \betao {\beta_{\omega}}
\def \epso {\epsilon_{\omega}}
\def \alfaks {\alpha_{\sst \rm K^*}}
\def \betaks {\beta_{\sst \rm K^*}}
\def \epsks {\epsilon_{\sst \rm K^*}}
\def \alfar {\alpha_{\rho}}
\def \betar {\beta_{\rho}}
\def \epsr {\epsilon_{\rho}}
\def \ckspcv {C_{\sst K^*}^{\sst \rm PC,V}}
\def \dkspcv {D_{\sst K^*}^{\sst \rm PC,V}}
\def \ckspv {C_{\sst K^*}^{\sst \rm PV}}
\def \dkspv {D_{\sst K^*}^{\sst \rm PV}}
\def \ckspct {C_{\sst K^*}^{\sst \rm PC,T}}
\def \dkspct {D_{\sst K^*}^{\sst \rm PC,T}}

\begin{document}
\preprint{
\vbox{
\hbox{ICCUB-11-134}
}}
\title{Current constraints on the EFT for the $\Lambda N \to NN$ transition}

\author{Axel P\'erez-Obiol}
\email[]{axel@ecm.ub.es}
\affiliation{Departament d'Estructura i Constituents de la Mat\`{e}ria,\\
Institut de Ci\`encies del Cosmos (ICC), \\
Universitat de Barcelona, Mart\'i Franqu\`es 1, E--08028, Spain}

\author{Assumpta Parre\~no}
\email[]{assum@ecm.ub.es}
\affiliation{Departament d'Estructura i Constituents de la Mat\`{e}ria,\\
Institut de Ci\`encies del Cosmos (ICC), \\
Universitat de Barcelona, Mart\'i Franqu\`es 1, E--08028, Spain}

\author{Bruno Juli\'a-D\'{\i}az}
\email[]{bruno@ecm.ub.es} 
\affiliation{Departament d'Estructura i Constituents de la Mat\`{e}ria,\\
Institut de Ci\`encies del Cosmos (ICC), \\
Universitat de Barcelona, Mart\'i Franqu\`es 1, E--08028, Spain}
\affiliation{ICFO-Institut de Ci\`encies Fot\`oniques, Parc Mediterrani de la Tecnologia, E-08860
Castelldefels (Barcelona), Spain}

\date{\today}

\begin{abstract}
The relation between the low energy constants appearing in the
effective field theory description of the $\Lambda N \to NN$
transition potential and the parameters of the one-meson-exchange model
previously developed are obtained. We extract the relative importance of the different exchange mechanisms included in the meson picture by means of a comparison to 
the corresponding operational structures appearing in the effective
approach. The ability of this procedure to obtain the weak baryon-baryon-meson couplings for a possible scalar exchange is also discussed.
\end{abstract}

\pacs{13.75.Ev, 21.80.+a, 25.80.Pw, 13.30.Eg}
\maketitle

\section{Introduction}
The use of effective field theory (EFT) approaches provides a
systematic way of handling nonperturbative strong
interaction physics. In particular, it is appealing for the
description of the short-distance physics of baryon-baryon
interactions. 

The EFT for the nonleptonic weak $|\Delta S|=1$
$\Lambda N$ interaction, which is the main responsible for the nonmesonic
decay of mostly all hypernuclei was first formulated in
Refs.~\cite{Jun} and \cite{PBH05}. While the authors in \cite{Jun} constructed
the effective theory by adding to the long-ranged 
one-pion-exchange mechanism (OPE) a four-fermion point interaction, coming from Lorentz 
four-vector currents, 
Ref.~\cite{PBH05} added the $K$- exchange mechanism (OKE) to the intermediate 
range of the interaction, as well as additional operational structures to the short 
range part of the transition
potential. These structures result when all possible operators compatible with the symmetries fulfilled by the weak $|\Delta S|=1$ $\Lambda N$ interaction are considered. The local operators
governing short distance dynamics in any EFT appear in the Lagrangian
multiplied by low energy constants (LECs), which have to be
determined by a fit to the available experimental data. Although
neither the amount nor the quality of hypernuclear weak decay data
is comparable with the wealth of
information available in the nonstrange sector, these data are
enough to fairly constrain the lowest order LECs. In order to
provide a higher order description of the weak 4-fermion
interaction, and therefore, a deeper understanding of the
fundamental dynamics involved, more and better data are needed, or
in their absence, a mapping to successful one-meson-exchange (OME)
models can be performed. Understanding these low energy constants in
terms of physical ingredients of the OME models, as masses, strong
form factor parameters and couplings of pseudoscalar and vector
mesons to baryons, is called resonance saturation~\cite{EKW93}
and it is the aim of the present manuscript.

The present work is partly motivated by the possible presence of
an isoscalar spin independent central transition operator in the weak
decay mechanism, and its relevant role in the prediction of some
hypernuclear decay observables~\cite{PBH05,SIO05,BM06}. 
This operational structure
would map a scalar $\sigma-$meson resonance in the traditional meson-exchange
picture. The fact that the $\sigma$ does not belong to the 
ground state meson octet has prevented its inclusion in many OME treatments of the weak
transition amplitude. Some works, however, have included the phenomenological
exchange of a correlated 2-$\pi$ (and/or 2-$\rho$ pair) state 
coupled to a scalar-isoscalar
channel, understood as a $\sigma$ resonance~\cite{IUM95,OJP01,BM05,CGPR07_sigma}, and
pointed out its relevance to determine the strength of some particular
transition amplitudes. The publication of new accurate data on hypernuclear decay observables during the last five years, makes it timely to revise the calculation of Ref.~\cite{PBH05}, and explore the 
feasibility of the EFT approach to constrain the weak baryon-baryon-sigma coupling constants.

To facilitate the reading of the present manuscript, and although the 
EFT formalism as well as
the OME one were developed and presented elsewhere,
we choose to include here an schematic overview of basics, 
together with the final relations governing the weak dynamics according to 
each one of the approaches.

\section{The meson exchange potential}
\label{MEP}

The $\Lambda$ hyperon decays in free space through the nonleptonic weak
decay modes $\Lambda \to n \pi^0$ and $\Lambda \to p \pi^-$, with an
approximate ratio
of 36:64. This mechanism is highly suppressed in the nuclear
medium, since the momentum of the nucleon in the final state is not large
enough to access unoccupied states above the Fermi energy level.
However, hypernuclear systems decay, precisely due to
the presence of surrounding nucleons, by means of single-, $\Gamma_{1N}=\Lambda N \to NN$, and
multi-nucleon induced decay mechanisms. Recently, the detection of two nucleons in coincidence in the final state~\cite{Outa05,Kang06,Kim06} has allowed a more reliable extraction not only of the 
total nonmesonic decay rate, but also of the ratio between the neutron induced process
($\Lambda n \to n n$) and the proton induced one ($\Lambda p \to n p$), $\Gamma_n/\Gamma_p$~\cite{GPR03,GPR04}. 
The analysis of the data points out that, in order to isolate the physical region where medium effects and multinucleon induced processes
are minimal, one needs to study the energy and angular correlated spectra for the particles detected in the final state, instead of looking at the absolute values for the partial and total decay rates.
Additionally, experiments performed with polarized hypernuclei, provide us with a measure of
the asymmetry in the angular distribution of protons in the final state, asymmetry that can be understood from the interference between
the parity-conserving (PC) and parity-violating (PV) weak amplitudes. The explicit expressions for the different decay rates, as well as the PV asymmetry, can be found in the original reference~\cite{PRB97}. 

\begin{figure}[t!]
\begin{center}
  \includegraphics[scale=0.3]{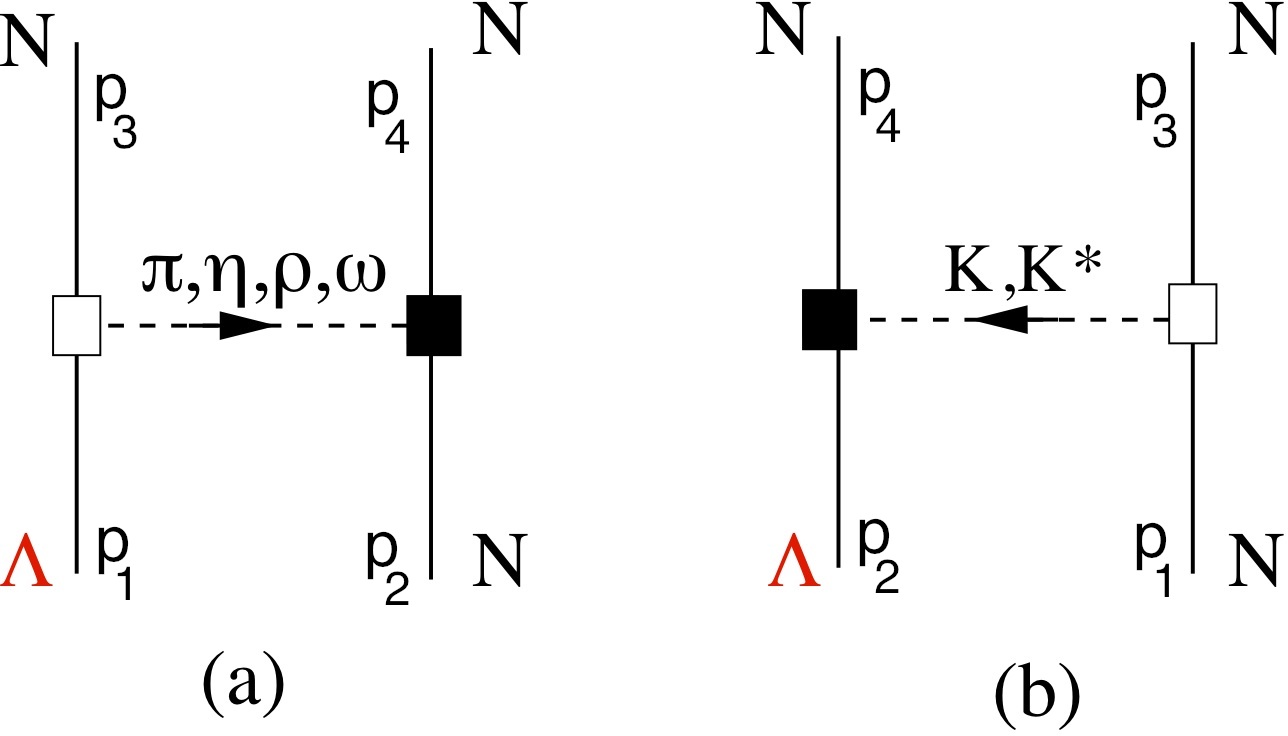}
\caption{
Non-strange (a) and strange (b) meson-exchange contributions to the
$\Lambda$N$\to$NN weak transition potential. A weak insertion is
indicated by the empty square, while a filled square stands for a strong interaction vertex.
}
\label{fig:amp}
\end{center}
\end{figure}

Traditionally, and in analogy with the strong $NN$ interaction, the one-nucleon induced decay mode, $\Lambda N \to NN$, has been described by a one-boson-exchange model,
according to which, a pion emitted at the weak $\Lambda N$ vertex
is absorbed by the $NN$ pair at the strong one. While mesons
other than the pion would be forbidden for the decay of the
$\Lambda$ particle in free space, there is no restriction for the
off-shell exchange of massive bosons. In the considered energy
domain, one needs to explicitly consider the exchange of the ground
state of pseudoscalar and vector meson octets. Higher energy physics
is parameterized through explicit cut-offs of $\approx 1$ GeV. The
momentum space transition potential will be therefore given by the
nonrelativistic limit of the appropriate Feynman amplitude
depicted in Figs.~\ref{fig:amp}(a) and~\ref{fig:amp}(b).

Using the strong and weak Hamiltonians given explicitely in
Appendix~\ref{appA}, the OPE potential reads:
\begin{equation}
{V_{\pi}} ({\vec q}\,) =
- G_F m_\pi^2 \,
\frac{g} {2 M_{\rm S}}
\left(
\hat{A} + \frac{\hat{B}}{2 M_{\rm W}}{\vec \sigma}_1 \, {\vec q} \,
\right)
\frac{{\vec \sigma}_2 \, {\vec q}\,} {{\vec q}^{\; 2}+\mu^2} \,
{\rm ,}
\label{eq:pion}
\end{equation}
where ${\vec q}=\vec{p}_1-\vec{p}_3$ is the momentum carried by the pion directed
towards the strong vertex, $g = g_{\sst {\rm NN} \pi}$ the strong
coupling constant for the NN$\pi$ vertex, $\mu$ the pion mass,
$M_{\rm S}$ ($M_{\rm W}$) the average of the baryon masses at the
strong (weak) vertex, and $\hat{A}=A_\pi \vec{\tau_1} \vec{\tau_2}$
and $\hat{B}=B_\pi \vec{\tau_1} \vec{\tau_2}$ the isospin operators
containing the weak parity-violating and parity-conserving coupling constants.

The $\eta$ and K exchanges, whose strong and weak vertices are again
explicitly given in Appendix~\ref{appA}, can be obtained from Eq.
(\ref{eq:pion}) by making the replacements:
$$
g \to  g_{\sst {\rm NN} \eta} \ , \, \,
\mu \to m_\eta  \ , \, \,
{\hat A} \to A_\eta \ , \, \,
{\hat B} \to B_\eta
$$
in the case of $\eta$-exchange, and
\begin{eqnarray}
g &\to& g_{\sst \Lambda {\rm N} K} \ , \, \,
\mu \to m_{\rm {\sst K}} \ , \, \, \nonumber \\
{\hat A} &\to &\left( \frac{ C^{\rm\sst{PV}}_{\rm\sst{K}}}{2} +
D^{\rm\sst{PV}}_{\rm\sst{K}} + \frac{
C^{\rm\sst{PV}}_{\rm\sst{K}}}{2} {\vec \tau}_1 {\vec \tau}_2
\,\right) 
\ , \, \,
\nonumber \\
{\hat B}& \to& \left( \frac{
C^{\rm\sst{PC}}_{\rm\sst{K}}}{2} +
D^{\rm\sst{PC}}_{\rm\sst{K}} + \frac{
C^{\rm\sst{PC}}_{\rm\sst{K}}}{2}
{\vec \tau}_1 \, {\vec \tau}_2 \right)
\end{eqnarray}
in the case of K-exchange. 

The short-range one-meson-exchange $\Lambda N$ interaction is
supplemented by the inclusion of more massive bosons, up to a mass
of around 1 GeV, the $\rho,~\omega$ and $K^*$ mesons. 
For the $\rho$-meson, for example, the non relativistic reduction of the pertinent Feynman amplitude,
computed using the vertices of Appendix~\ref{appA}, gives the
following transition potential:
\begin{eqnarray}
{V_{\rho}}({\vec q}\,)  &=&
 \left[ F_1 {\hat \alpha} - \frac{({\hat \alpha} + {\hat \beta} )
 ( F_1 + F_2 )} {4M_{\rm S} M_{\rm W}}
({\vec \sigma}_1 \, \times {\vec q} \,)
({\vec \sigma}_2 \, \times {\vec q} \,) \right. \nonumber \\
& &
\left. - \im \frac{{\hat \varepsilon} ( F_1 + F_2 )} {2M_{\rm S}}
({\vec \sigma}_1 \, \times {\vec \sigma}_2 \, ) {\vec q} \,\right]
\frac{G_F m_\pi^2}{{\vec q}^{\; 2} + \mu^2} \ , \label{eq:rhopot}
\end{eqnarray}
with $\mu = m_\rho$, $F_1 = g^{\rm {\sst V}}_{\rm {\sst NN} \rho}$,
$F_2 = g^{\rm {\sst T}}_{\rm {\sst  NN} \rho}$ and where the
operators ${\hat \alpha}$, ${\hat \beta}$ and ${\hat \varepsilon}$,
defined by:
$$
{\hat \alpha} = \alpha_\rho \,\, {\vec \tau}_1 \,
{\vec \tau}_2  \ , \,\,\,
{\hat \beta} = \beta_\rho \,\, {\vec \tau}_1
{\vec \tau}_2  \ , \,\,\, {\rm and} \,\,
{\hat \varepsilon} = \varepsilon_\rho \,\, {\vec \tau}_1 \,
{\vec \tau}_2 \ ,
$$
contain the isospin structure in addition to the weak coupling
constants.

The nonrelativistic potential can be obtained from the
general expression given in Eq. (\ref{eq:rhopot}) by
making the following replacements:
\begin{eqnarray}
&&\mu \to m_\omega \ , \,\,\,
F_1 \to g^{\rm {\sst V}}_{\rm {\sst NN} \omega} \ , \,\,\,
F_2 \to g^{\rm {\sst T}}_{\rm {\sst NN} \omega} \ , \nonumber \\
&&{\hat \alpha} \to \alpha_\omega  \ , \,\,\,
{\hat \beta} \to \beta_\omega  \ , \,\,\,
{\hat \varepsilon} \to \varepsilon_\omega
\end{eqnarray}
in the case of $\omega$-exchange, and
\begin{eqnarray}
\mu &\to& m_{\rm {\sst K^*}} \ , \,\,
F_1 \to g^{\rm \sst{V}}_{\rm {\sst \Lambda N K^*}} \ , \,\,
F_2 \to g^{\rm \sst{T}}_{\rm {\sst \Lambda N K^*}} \nonumber \\
{\hat \alpha} &\to& \frac{
C^{\rm \sst{PC, V}}_{\rm \sst{K^*}}} {2} +
D^{\rm \sst{PC, V}}_{\rm \sst{K^*}} + \frac{
C^{\rm \sst{PC, V}}_{\rm \sst{K^*}}} {2}
{\vec \tau}_1 \, {\vec \tau}_2  \nonumber \\
{\hat \beta} &\to& \frac{
C^{\rm \sst{PC, T}}_{\rm \sst{K^*}}} {2} +
D^{\rm \sst{PC, T}}_{\rm \sst{K^*}} + \frac{
C^{\rm \sst{PC, T}}_{\rm \sst{K^*}}} {2}
{\vec \tau}_1 \, {\vec \tau}_2 \nonumber \\
{\hat \varepsilon} &\to& \left( \frac { C^{\rm \sst{PV}}_{\rm
\sst{K^*}}} {2} + D^{\rm \sst{PV}}_{\rm \sst{K^*}} + \frac{ C^{\rm
\sst{PV}}_{\rm \sst{K^*}}} {2} {\vec \tau}_1 \, {\vec \tau}_2
\,\right) 
\end{eqnarray}
for the exchange of a K$^*$-meson.
Note that the K$^*$ weak vertex has the same structure as the K
one, the only difference being the parity-conserving contribution
which has two terms, related to the vector and tensor couplings.

Due to the lack of enough phase space to produce the desired
decay vertex, the baryon-baryon-meson couplings 
for mesons heavier than the pion are not available experimentally.
To fix such couplings one uses SU(3) flavor (SU(6) spin-flavor) symmetry to relate 
the unknown couplings involving pseudoscalar (vector) mesons to the pionic decay vertex.
For the strong vertices we use the values given by 
the Nijmegen Soft-Core f~\cite{nij99} and the J\"ulich B~\cite{JB} models, which also rely on
the same symmetries. This choice generates a model dependency in our approach, which also
propagates to the weak couplings through the pole model~\cite{DFHT96}
used to evaluate the weak PC baryon-baryon-meson constants. In order to be consistent, 
we use the same strong potential models to derive the scattering (T-matrix) NN wave functions in the
final state~\cite{PR02}.

To regularize the potentials at higher energies we include a form factor at each vertex of the OME diagram. The form of this form factor depends on the strong interaction model we are considering. In the case of the J\"ulich B model we use a monopole form factor, $F(\vec{q})=\left({\Lambda_i^2-\mu_i^2\over \Lambda_i^2 + \vec{q} \, ^2}\right)$, at each vertex, while for the Nijmegen SC97 models, we use a modified monopole version~\cite{PR02}, 
$F(\vec{q})=\left({\Lambda_i^2\over \Lambda_i^2 + \vec{q} \, ^2}\right)$.
In both cases, the value of the cut-off, $\Lambda_i$, depends on the meson exchanged (with mass $\mu_i$). 
The full set of meson-exchange parameters employed here is
given in Tables~\ref{tabpar} and \ref{tabparj}.

\begin{table}
\centering
\begin{tabular}{c|l|l|l|l}
M   & \multicolumn{1}{c|}{Strong c.c.} & \multicolumn{2}{c|}{Weak c.c.} &
$~~\Lambda_i$ \\
        &     & \multicolumn{1}{c}{PC} & \multicolumn{1}{c|}{PV} &
        \mbox{(GeV)} \\
\hline
\hline
 $\pi$  & $g_{\sst {\rm NN} \sst\pi}$ = 13.16 & $B_\pi$=$-$7.15 & 
$A_\pi$=1.05 & 1.750  \\
\hline
 $\eta$ & $g_{\sst {\rm NN} \sst \eta}$ = 6.42 & $B_\eta$=$-$11.9 & 
$A_\eta$=1.80 & 1.750 \\
\hline
K       & $g_{\sst \Lambda\rm NK}$ = $-$17.66  &
$C_{\sst \rm K}^{\sst {\rm PC}}$=$-$23.70 &
$C_{\sst \rm K}^{\sst {\rm PV}}$=0.76 & 1.789 \\
    & $g_{\sst \rm N \Sigma K}$ = 5.38 & 
$D_{\sst \rm K}^{\sst {\rm PC}}$=8.33 & $D_{\sst \rm K}^{\sst {\rm PV}}$=2.09
& \\
\hline
$\rho$ & $g_{\sst {\rm NN} \rho}^{\sst {\rm V}}$ = 2.97 & 
$\alpha_\rho$=$-$3.29 & $\epsilon_\rho$=1.09
& 1.232 \\
       & $g_{\sst {\rm NN} \rho}^{\sst {\rm T}}$ = 12.52  & 
$\beta_\rho$=$-$6.74&  &  \\
\hline
$\omega$ & $g_{\sst {\rm NN} \omega}^V$ = 10.36 & 
$\alpha_\omega$=$-$0.17  &
$\epsilon_\omega$= $-$1.33 & 1.310 \\
      & $g_{\sst {\rm NN} \omega}^T$ = 4.195 &$\beta_\omega$=$-$7.43  &  & \\
\hline
K$^*$ & $g_{\sst \Lambda \rm N K^*}^{\sst \rm V}$ =$-$6.105 &
$C^{\sst  {\rm PC,V}}_{\sst \rm K^*}$=$-$4.02 &
$C^{\sst  {\rm PV}}_{\sst \rm K^*}$=$-$4.48 & 1.649
 \\
     & $g_{\sst \Lambda \rm N K^*}^{\sst \rm T}$ = $-$14.85 &
$C^{\sst {\rm PC,T}}_{\sst \rm  K^*}$=$-$19.54 &
& \\
   &                                      &  
$D^{\sst {\rm PC,V}}_{\sst \rm K^*}$=$-$5.46
   &
  $D^{\sst {\rm PV}}_{\sst \rm K^*}$=0.60 & \\
   &                                      &  
$D^{\sst {\rm PC,T}}_{\sst \rm K^*}$=6.23 &  &
 \\
\end{tabular}
\caption{Nijmegen (NSC97f) meson exchange parameters used in the present work.
The weak couplings are in units of $G_F {m_\pi}^2 = 2.21 \times 10^{-7} $.
\label{tabpar}}
\end{table}

\begin{table}
\centering
\begin{tabular}{c|l|l|l|l}
M   & \multicolumn{1}{c|}{Strong c.c.} & \multicolumn{2}{c|}{Weak c.c.} &
$~~\Lambda_i$ \\
        &     & \multicolumn{1}{c}{PC} & \multicolumn{1}{c|}{PV} &
        \mbox{(GeV)} \\
\hline
\hline
 $\pi$  & $g_{\sst {\rm NN} \sst\pi}$ = 13.45 & $B_\pi$=$-$7.15 & 
$A_\pi$=1.05 & 1.300  \\
\hline
 $\eta$ & $g_{\sst {\rm NN} \sst \eta}$ = 0 & $B_\eta$=0 & 
$A_\eta$=1.80 & 1.300 \\
\hline
K       & $g_{\sst \Lambda\rm NK}$ = $-$13.48  &
$C_{\sst \rm K}^{\sst {\rm PC}}$=$-$17.67 &
$C_{\sst \rm K}^{\sst {\rm PV}}$=0.76 & 1.200 \\
    & $g_{\sst \rm N \Sigma K}$ = 3.55 & 
$D_{\sst \rm K}^{\sst {\rm PC}}$=5.50 & $D_{\sst \rm K}^{\sst {\rm PV}}$=2.09
& \\
\hline
$\rho$ & $g_{\sst {\rm NN} \rho}^{\sst {\rm V}}$ = 3.25 & 
$\alpha_\rho$=$-$3.60 & $\epsilon_\rho$=1.09
& 1.400 \\
       & $g_{\sst {\rm NN} \rho}^{\sst {\rm T}}$ = 19.82  & 
$\beta_\rho$=$-$9.55&  &  \\
\hline
$\omega$ & $g_{\sst {\rm NN} \omega}^V$ = 15.85 & 
$\alpha_\omega$=$-$5.85  &
$\epsilon_\omega$= $-$1.33 & 1.500 \\
      & $g_{\sst {\rm NN} \omega}^T$ = 0 &$\beta_\omega$=$-$10.96  &  & \\
\hline
K$^*$ & $g_{\sst \Lambda \rm N K^*}^{\sst \rm V}$ =$-$5.63 &
$C^{\sst  {\rm PC,V}}_{\sst \rm K^*}$=$-$3.71 &
$C^{\sst  {\rm PV}}_{\sst \rm K^*}$=$-$4.48 & 2.200
 \\
     & $g_{\sst \Lambda \rm N K^*}^{\sst \rm T}$ = $-$18.34 &
$C^{\sst {\rm PC,T}}_{\sst \rm  K^*}$=$-$26.38 &
& \\
   &                                      &  
$D^{\sst {\rm PC,V}}_{\sst \rm K^*}$=$-$5.03
   &
  $D^{\sst {\rm PV}}_{\sst \rm K^*}$=0.60 & \\
   &                                      &  
$D^{\sst {\rm PC,T}}_{\sst \rm K^*}$=12.18 &  &
 \\
\end{tabular}
\caption{Same as Table~\ref{tabpar} but for the J\"ulich B model.
\label{tabparj}}
\end{table}

\section{The Effective Field Theory approach}

To a given order in the EFT approach, the weak nonleptonic $\Lambda
N \to NN$ interaction is built by adding to the $\pi$ and $K$
exchange mechanisms a series of local terms with increasing
dimension ({\it i.e.} increasing number of derivatives) and
compatible with chiral symmetry, Lorentz invariance and the
applicable discrete symmetries.

\begin{figure}
\vspace*{0.5cm}
\includegraphics[scale=0.2]{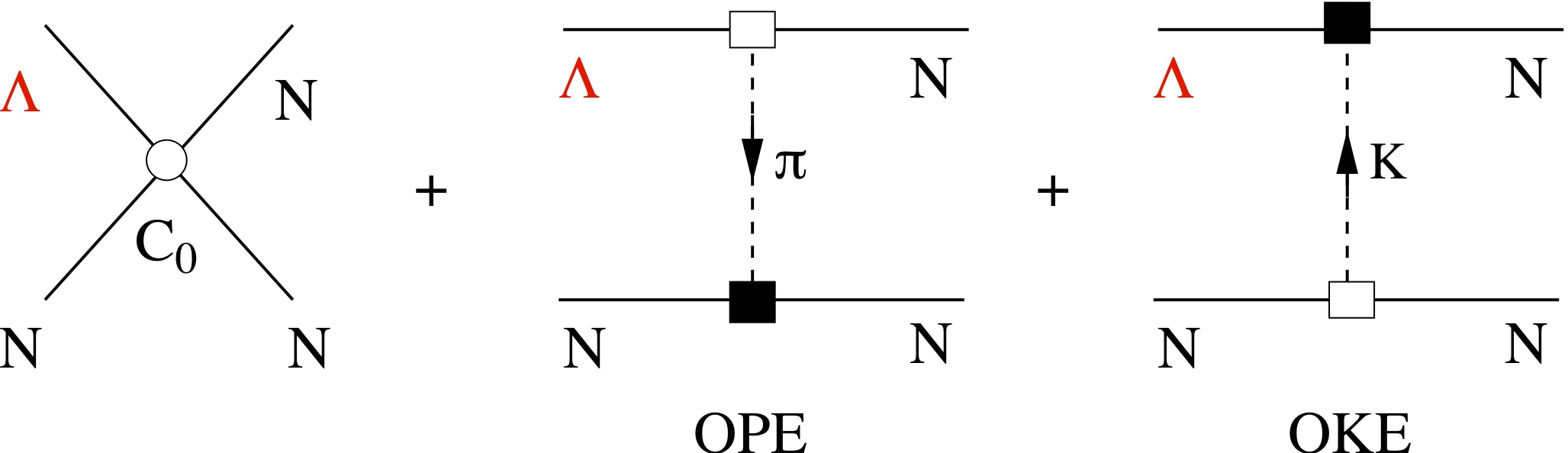}
\caption{Lowest order contribution to the weak $\Lambda N \to NN$ diagram. 
Empty symbols represent weak vertices while solid ones 
represent strong vertices. A circle stands for a contact non derivative
operator and a square for an insertion of a derivative operator.}
\label{fig:lo}
\end{figure}

Therefore, the leading order (LO) contribution will contain, apart from
the OPE and OKE diagrams, contact operators with no derivatives acting on the four-baryon vertex.
The inclusion of the long ranged $\pi$-exchange mechanism is
justified by the high value of the momentum transfer in the weak
reaction, $|{\vec q}\,| \sim 400$ MeV, a consequence of the
difference between the $\Lambda$ and nucleon masses in the initial
state. The same argument holds for the explicit inclusion of the $K$
meson, supported also by chiral symmetry.
From the diagramatical point of view the LO contribution to the
potential is given by Fig.~\ref{fig:lo}.

\ignore{
\begin{figure}
\includegraphics[scale=0.2]{nlo.jpg}
\caption{Next-to-leading order (leading order PV) contribution to the
$\Lambda N \to NN$ diagram. Same convention for the drawing of the vertices 
as in Fig.~\ref{fig:lo}}
\label{fig:nlo}
\end{figure}

\begin{figure}
\includegraphics[scale=0.3]{nnlo.jpg}
\caption{Next-to-next to leading order (next-to-leading PC)
contributions to the $\Lambda N \to NN$ diagram.}
\label{fig:nnlo}
\end{figure}
}
One may, equivalently, proceed to chirally expand the vertices entering the
$\Lambda N \to NN$ transition, and use a phenomenological approach to
account for the strong interaction between the baryons involved in the process.
Those vertices are nothing else but combinations of the five Dirac bilinear
covariants; $1$, $\gamma^5$, $\gamma^\mu$, $\gamma^\mu\gamma^5$ and
$\frac{i\sigma^{\mu\nu}q_\nu}{2M}$; where $\sigma^{\mu\nu}=\frac
i2[\gamma^\mu,\gamma^\nu]$, $M$ the mass of the baryon and $q_\nu$
the transferred momentum.
Since the relativistic form of these bilinears encodes all the
orders in a momentum expansion,
it is their chiral expansion which will better allow the power counting by
comparing non relativistic terms of size 1, p/M, etc.
In order to avoid formal inconsistencies from the chiral point of view,
we rely directly on the terms which enter at each order given the 
symmetries fulfilled by the weak $|\Delta S|=1$ transition.
\begin{table}[tb]
\begin{tabular}{lccccccccc}
{\rm partial \,\, wave} & & & {\rm operator} & & & {\rm size} & & I \\
\hline
$a: ^1S_0 \to ^1S_0$ & & &${\hat 1}, \, {\vec \sigma_1} {\vec \sigma_2}$
&& & $1$ & & & $1$ \\
$b: ^1S_0 \to ^3P_0$ & & &$({\vec \sigma_1} - {\vec \sigma_2}) {\vec q} {\rm ,}
\,\, ({\vec \sigma_1} \times {\vec \sigma_2}) {\vec q}$ & & & $q/M_N$ & & &  $1$
 \\
$c: ^3S_1 \to ^3S_1$ & & &${\hat 1}, {\vec \sigma_1} {\vec \sigma_2}$ & & &$1$ &
 &&  0 \\
$d: ^3S_1 \to ^1P_1$ & & &$({\vec \sigma_1} - {\vec \sigma_2}) {\vec q} {\rm ,}
\,\, ({\vec \sigma_1} \times {\vec \sigma_2}) {\vec q} $ & &&  $q/M_N$ & && $0$
\\
$e: ^3S_1 \to ^3P_1$ & & &$({\vec \sigma_1} + {\vec \sigma_2}) {\vec q}$
& & & $q/M_N$ && &  $1$ \\
$f: ^3S_1 \to ^3D_1$ & & &$({\vec \sigma_1} \times {\vec q}) ({\vec \sigma_2}
\times {\vec q})$ & & &$q^2/{M_N}^2$ & & & $0$ \\
\end{tabular}
\caption{$\Lambda N \to NN$ transitions for an initial $\Lambda N$ relative 
$S-$wave state.\label{tab:pw}}
\end{table}
All these possible transitions are shown in Table~\ref{tab:pw} for
an initial $S-$wave $\Lambda-N$ state,
where, the model independent leading order operators in momentum space responsible for
the transitions are listed (we are
assuming that $|{\vec p}_1 - {\vec p}_2|$ is small enough to disregard
higher powers of the derivative operators $ {\vec p}_1 - {\vec p}_2
$). Organizing all these contributions in increasing size
operators, we obtain the most general Lorentz invariant potential,
with no derivatives in the fields, for the four-fermion (4P)
interaction in momentum space up to ${\cal O} (q^2/M^2)$ order (in
units of $G_F = 1.166 \times 10^{-11}$ MeV$^{-2}$):
\begin{eqnarray}
V_{4P} ({\vec q} \, ) &=&
C_0^0 + C_0^1 \; {\vec \sigma}_1 
{\vec \sigma}_2 \label{eq:vnnlo}
\\
&+& C_1^0 \; \displaystyle\frac{{\vec \sigma}_1 
{\vec q}}
{2 {\overline M}}
+ \,C_1^1 \; \displaystyle\frac{{\vec \sigma}_2
{\vec q}}{2 M}
+ {\rm i} \, C_1^2 \; \displaystyle\frac{({\vec \sigma}_1 \times {\vec \sigma}_2)
\; 
{\vec q}}{2 \tilde{M}} \nonumber
\\
&+& C_2^0 \; \displaystyle\frac{{\vec \sigma}_1 
{\vec q} \;
{\vec \sigma}_2 
{\vec q}}{4 M {\overline M}} +
C_2^1 \; \displaystyle\frac{{\vec \sigma}_1
{\vec \sigma}_2 \; {\vec q}^{\; 2}}
{4 M {\overline M}} +
C_2^2 \; \displaystyle\frac{{\vec q}^{\,2}}{4 M \tilde{M}} \ {\rm ,}
\nonumber
\end{eqnarray}
\noindent where $M$ is the nucleon mass, 
${\overline M} = (M + M_\Lambda)/2$,
${\tilde M} = (3 M + M_\Lambda)/4$ (with $M_\Lambda$ the $\Lambda$ mass) and $C_i^j$ is the j$th$ low
energy coefficient at i$th$ order. To derive the previous
expression, we have used the relation $({\vec \sigma}_1 \,
\times {\vec q} \,) ({\vec \sigma}_2 \, \times {\vec q}\,) = ({\vec
\sigma}_1 \, {\vec \sigma}_2)\: {\vec q}^{\ 2} - ({\vec \sigma}_1 \,
{\vec q} \,) ({\vec \sigma}_2 \,{\vec q}\,) \ {\rm }$. 
Notice that, in principle, one could write, at next-to-next-to leading order, NNLO, another set of eight operators
containing the isospin structure $\vec\tau_1\cdot\vec\tau_2$. However,
once one imposes that the final two-nucleon state must be antisymmetric, the number of structures in the effective potential is reduced to half the original, leaving to only eight independent operators.

\ignore{
{\bf {\it At Lagrangian level one can relate the direct diagrams
with the crossed ones using the Fierz identities. Equivalently, an
effective potential with sixteen terms (understood as coming from only
direct diagrams), can be antisymmetrized imposing the
$L+S+T=odd$ rule (see for example Refs.~\cite{we97} and \cite{ep01}). We
have chosen the previous base of eight operators following Ref. ~\cite{ep01}, but many choices are possible.}}}

The relation between the LO constants appearing in Eq.~(\ref{eq:vnnlo})
and the ones in the non-antisymmetrized potential,
\begin{eqnarray}
V_{4P}^{LO'}({\vec q} \,
)&=&C_{0~sc}^0+C_{0~vec}^0\;\vec\tau_1\vec\tau_2
\nonumber\\\nonumber
&+&
C_{0~sc}^1\,\vec\sigma_1\vec\sigma_2+
C_{0~vec}^1\,\vec\sigma_1\vec\sigma_2\;\vec\tau_1\vec\tau_2
\label{eq:vlo2}
\end{eqnarray}
is the following (see Ref. ~\cite{epth}):
\begin{eqnarray}
C_0^0&=&C_{0~sc}^0-2\;C_{0~vec}^0-3\,C_{0~vec}^1
\nonumber\\
C_0^1&=&C_{0~sc}^1-C_{0~vec}^0 \ {\rm .}
\label{eq:base}
\end{eqnarray}

\ignore{
With respect to the isospin part of the 4-fermion interaction, we allow
for both, $\Delta I=1/2$ and $\Delta I=3/2$ transitions \footnote {
Matrix elements of the $\Delta I=1/2$ ($\Delta I=3/2$) operator can
be easily included by assuming the $\Lambda$ to behave like an
isospin$\mid 1/2 \ -1/2 \rangle$ ($\mid 3/2 \ -1/2 \rangle$) state
and introducing an isospin dependence in the $\Delta$I=1/2
($\Delta$I=3/2) transition potential of the type \mbox{${\vec \tau}
{\vec \tau} \, $}(\mbox{${\vec \tau}_{3/2} {\vec \tau} \, $}),
where ${\vec \tau}$ (${\vec \tau}_{3/2}$) is the $1/2 \rightarrow
1/2$ ($1/2 \rightarrow 3/2$) isospin transition operator. The
spherical components of the $\Delta I=1/2$ and $\Delta I=3/2$
operators have matrix elements~\protect\cite{PRBM98} $ \, \langle
1/2\ m^\prime \mid \tau_{1/2}^{(i)} \mid 1/2\ m \rangle =\langle
1/2\
 m\ 1\ i  \mid 1/2\ m^\prime \rangle$ and
$\langle 3/2\ m^\prime \mid \tau_{3/2}^{(i)} \mid 1/2\ m \rangle =\langle
1/2\ m\ 1\ i  \mid 3/2\ m^\prime \rangle$, respectively, 
with $i=\pm 1,0$. Matrix elements of the 
${\hat {\cal O}}= C_{\sst \rm IS} \, {\hat 1} + C_{\sst \rm IV} \, {\vec \tau_1} {\vec \tau}_2
+ C_{\sst 3/2} \, {\vec \tau}_{3/2} {\vec \tau}_2 $ operator
read:
\mbox{
$
< n n | {\hat {\cal O}} | \Lambda n> =C_{\sst \rm IS} + C_{\sst \rm IV} - \sqrt{\frac{2}{3}} 
C_{\sst 3/2} \
{\rm ,}  < n p | {\hat {\cal O}} | \Lambda p> =
$
}
\mbox{
$ C_{\sst \rm IS} - C_{\sst \rm IV} + \sqrt{\frac{2}{3}} C_{3/2}
$}
,
\mbox{
$
< p n | {\hat {\cal O}} | \Lambda p> = 2 \, C_{IV} + 
\sqrt{\frac{2}{3}} C_{3/2} {\rm .} $
}
}
.
}

From the former derivation, it is clear that the form of the contact
terms is model independent. The LEC's represent the short distance
contributions and their size depends on how the theory is
formulated, and more specifically upon the chiral order we are
working. The low energy parameters which size the relative
contribution of the contact 4-fermion operators are fitted to the
known weak decay observables discussed in section~\ref{MEP}.

\ignore{At this point, we must say that the hypernuclear decay observables
available constrain only
enough to fit the effective potential at LO. }
\ignore{In order to regularize
the effective potential we have made use of Nijmegen (Ref.~\cite{nij})
and Jülich (Ref.~\cite{nij}) cut-offs. Table ~\ref{tab:lecs} shows}

\section{Relations between the OME potentials and the EFT}

To relate the meson-exchange constants to the LECs in the effective
$\Lambda N \to NN$ potential, we perform a low-momentum expansion of the
various (regularized) meson-exchange potentials other than the pion and the kaon, since these
two are explicitely included in both, the OME and the EFT approaches.
 This procedure leads to a
series of contact terms organized by their increasing dimension,
{\it i.e.} with increasing powers of momenta, an appropriate form 
to compare with the EFT potential.
Therefore, one can write these terms up to ${\cal O} ({\vec q \,}^2/M^2)$ order 
(in units of $G_F = 1.166 \times 10^{-11}$ MeV$^{-2}$) as :

\begin{eqnarray}
V^{\rm LO}_{\rm OME} (\vec{q} \,) &=&
     \left[ \,
     \frac{ \vcks}{{m_{\sst K^*}}^2 }\left(\frac{\ckspcv}{2}+\dkspcv\right)
    \, + \,
     \frac{ \vco \, \alfao}{{m_{\omega}}^2 } \right .
      \nonumber \\ 
     &+& \, \left .
     \left(
       \frac{\vcks\,\ckspcv}{2{m_{\sst K^*}}^2}
      \, + \,
       \frac{ \vcr \, \alfar}{{m_{\rho}}^2 }
       \right)\vec\tau_1\cdot\vec\tau_2
       \, \right] \, {m_{\pi}}^2 \,\, {\rm ,}
       \nonumber \\
\end{eqnarray}

\begin{widetext}
\begin{eqnarray}
V^{\rm NLO}_{\rm OME} (\vec{q} \,) &=&
\frac{-{m_{\pi}}^2}{2 M} 
      \frac{\, A_\eta \, g_{\sst {\rm NN} \eta}}
{{m_\eta}^2 } \, \vec{\sigma}_2 \vec{q}
     \,-\,
     \frac{{m_{\pi}}^2}{2 M} \left[
      \left(
      \frac{{\rm i} \, ( \vcks + \tcks ) \,(\frac{\ckspv}{2}+\dkspv) \, {m_{\pi}}^2 \, }
           { {m_{\sst K^*}}^2 }  \, + \,
      \frac{{\rm i} ( \vco + \tco ) \epso \, {m_{\pi}}^2 \, }
       { {m_{\omega}}^2 }
       \right)
       \right . \nonumber\\
    &+& \left.
      \left(
      \frac{{\rm i} \, ( \vcks + \tcks ) \,\ckspv \, {m_{\pi}}^2 \, }
           { {2m_{\sst K^*}}^2 }  \, + \,
      \frac{{\rm i} \, ( \vcr + \tcr ) \epsr \, {m_{\pi}}^2 \, }
       { {m_{\rho}}^2 }
      \right)\, \vec\tau_1\,\vec\tau_2
    \right]\, (\vec{\sigma}_1 \times \vec{\sigma}_2)\, \vec{q} \,\, {\rm ,} \nonumber\\
V^{\rm NNLO}_{\rm OME} (\vec{q} \,)\nonumber &=&
         \frac{{m_\pi}^2}{4M\overline{M}}\left[\left(
         \frac{\ckspcv}{2}+\dkspcv+\frac{\ckspct}{2}+\dkspct\right)\frac{\vcks+\tcks}{{m_{\sst K*}}^2}
         \,+\,
         \frac{\left(\alfao+\betao\right)\left(\vco+\tco\right)}{{m_\omega}^2}
          \right .
          \\\nonumber 
         &+& \left .
        \left(\frac{\left(\ckspcv+\ckspct\right)\left(\vcks+\tcks\right)}{2{m_{\sst K*}}^2}
         \,+\,
         \frac{\left(\alfar+\betar\right)\left(\vcr+\tcr\right)}{{m_\rho}^2}
         \right)\vec\tau_1\,\vec\tau_2\right]
         \,\left(\vec\sigma_1\,\vec{q}\,\,\vec\sigma_2\,\vec{q}
         -\vec\sigma_1\,\vec\sigma_2\,\vec q^{\,\,2}\right)
         \\\nonumber
         &-&\frac{{m_\pi}^2}{4M\overline{M}}
         \frac{B_\eta\,\ce}{{m_\eta}^2}\vec\sigma_1\,\vec{q}\,\,\vec\sigma_2\,\vec{q}
         \,-\,2{m_\pi}^2\left[\frac{\vcks\,\left(\Lambda^2+{m_{\sst
                 K^*}}^2\right)
             \left(\frac{\ckspcv}{2}+\dkspcv\right)}{{m_{\sst
                 K^*}}^4\Lambda^2}
           \,+\,\frac{\vco\alfao\left(\Lambda^2+{m_\omega}^2\right)}{{m_\omega}^4\Lambda^4}\right
           .
           \\\nonumber &+& \left .
           \left(\frac{\vcks\,\left(\Lambda^2+{m_{\sst
                 K^*}}^2\right)
             \,\ckspcv}{{2m_{\sst
                 K^*}}^4\Lambda^2}
           \,+\,\frac{\vcr\alfar\left(\Lambda^2+{m_\rho}^2\right)}{{m_\rho}^4\Lambda^4}\right)
          \, \vec\tau_1\,\vec\tau_2\right]\vec q^{\,2} \,\, {\rm .}
         \end{eqnarray}

\end{widetext}
We have chosen to show the explicit expressions of the LECs in terms of meson-exchange parameters in the Appendix~\ref{appB}. Here we only quote the relations at LO. In order to compare these expressions with the 4P potential of
Eq.~(\ref{eq:vnnlo}) we need to use the same base
of operators. Eq.(\ref{eq:base}) allows us to obtain the LO coefficients
in the $\hat{1}$, ${\vec \sigma}_1\cdot{\vec \sigma}_2$ base, 
\begin{eqnarray}
C_0^0&=&\nonumber
     \left[ \,
     \frac{ \vcks}{{m_{\sst K^*}}^2 }\left(\frac{\ckspcv}{2}+\dkspcv\right)
    \, + \,
     \frac{ \vco \, \alfao}{{m_{\omega}}^2 } 
\right .\\
\label{lecc00}
&-&\left .
       \frac{\vcks\,\ckspcv}{2{m_{\sst K^*}}^2}
      \, - \,
       \frac{ \vcr \, \alfar}{{m_{\rho}}^2 }
\right]\, {m_{\pi}}^2 \,\, {\rm ,}
\\
C_0^1&=&
\left[
-       \frac{\vcks\,\ckspcv}{2{m_{\sst K^*}}^2}
      \, - \,
       \frac{ \vcr \, \alfar}{{m_{\rho}}^2 } 
\right]\, {m_{\pi}}^2 \,\, {\rm .}
\label{lecc01}
\end{eqnarray}

\ignore{
Comparing these expressions to the 4P potential of
Eq.~(\ref{eq:vnnlo}), we can extract expressions for the LECs in
terms of meson-exchange parameters. The explicit formulae are given
in Appendix~\ref{appB}.}

In Table~\ref{tab:lecs} we show the results for the LECs obtained
within both formalisms. On the one hand, we quote the values for the
coefficients obtained from Eqs.~(\ref{lecc00}) and~(\ref{lecc01}) (left column, under the label {\it OME expansion}).  The numerical values for the constants in front of the spin-isospin operators
have been obtained for each strong interaction model, and Eq.(\ref{eq:base})
has been used to write the LECs in the antisymmetric base of operators. 
On the other hand, we show the values obtained from a fit of our EFT to reproduce the experimental data described in section \ref{MEP} (right column,
under the label {\it LO calculation}). We note that it is enough to consider the
LO EFT ({\it i.e.} just two constants) to obtain a reasonable fit to the data.
Notice that the values derived from the OME approach do not arise from any fit to the observables but from symmetry considerations
together with studies of the strong baryon-baryon interaction. Their
errors are estimated considering an uncertainity in the couplings of
$\pm30\%$. 
\ignore{The
comparison between both results shows a clear disagreement between the
OME approach and the EFT approach. }

\begin{figure}[th]
\begin{center}
\hspace*{-0.5cm}
\includegraphics[scale=0.40]{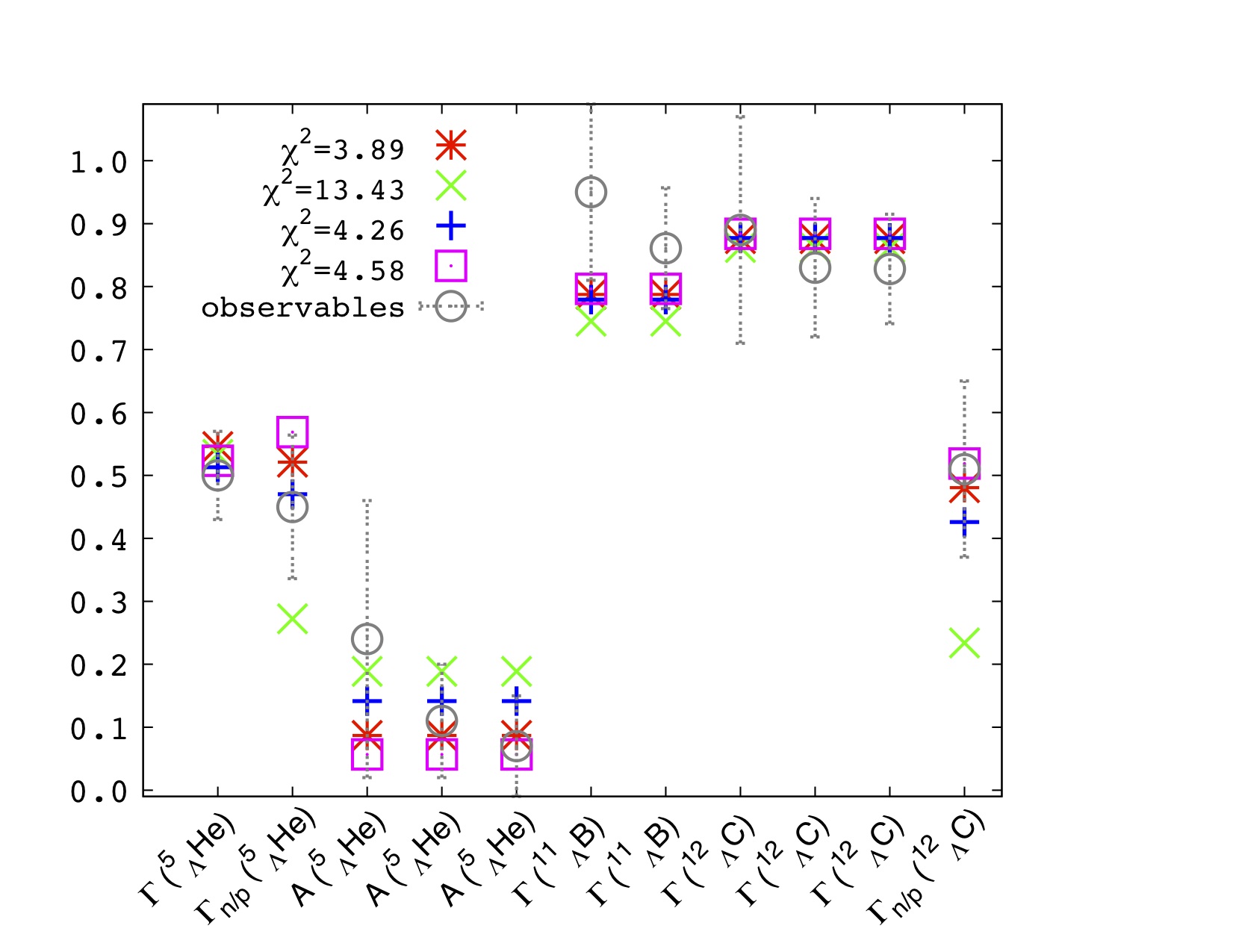}
\caption{Hypernuclear decay observables (total and partial
  decay rates and asymmetry for $^\Lambda_5\rm He$, $^\Lambda_{11}\rm
  B$ and $^\Lambda_{12}\rm C$), including their error bars and their
  fitted values.  The total decay rates are in units of the 
  $\Lambda$ decay rate in free space ($\Gamma_\Lambda=3.8\times10^9$s$^{-1}$). All the quantities are adimensional.}
\label{fits}
\end{center}
\end{figure}

\begin{figure}[ht!]
\begin{center}
\hspace*{-1cm}
\includegraphics[scale=0.41]{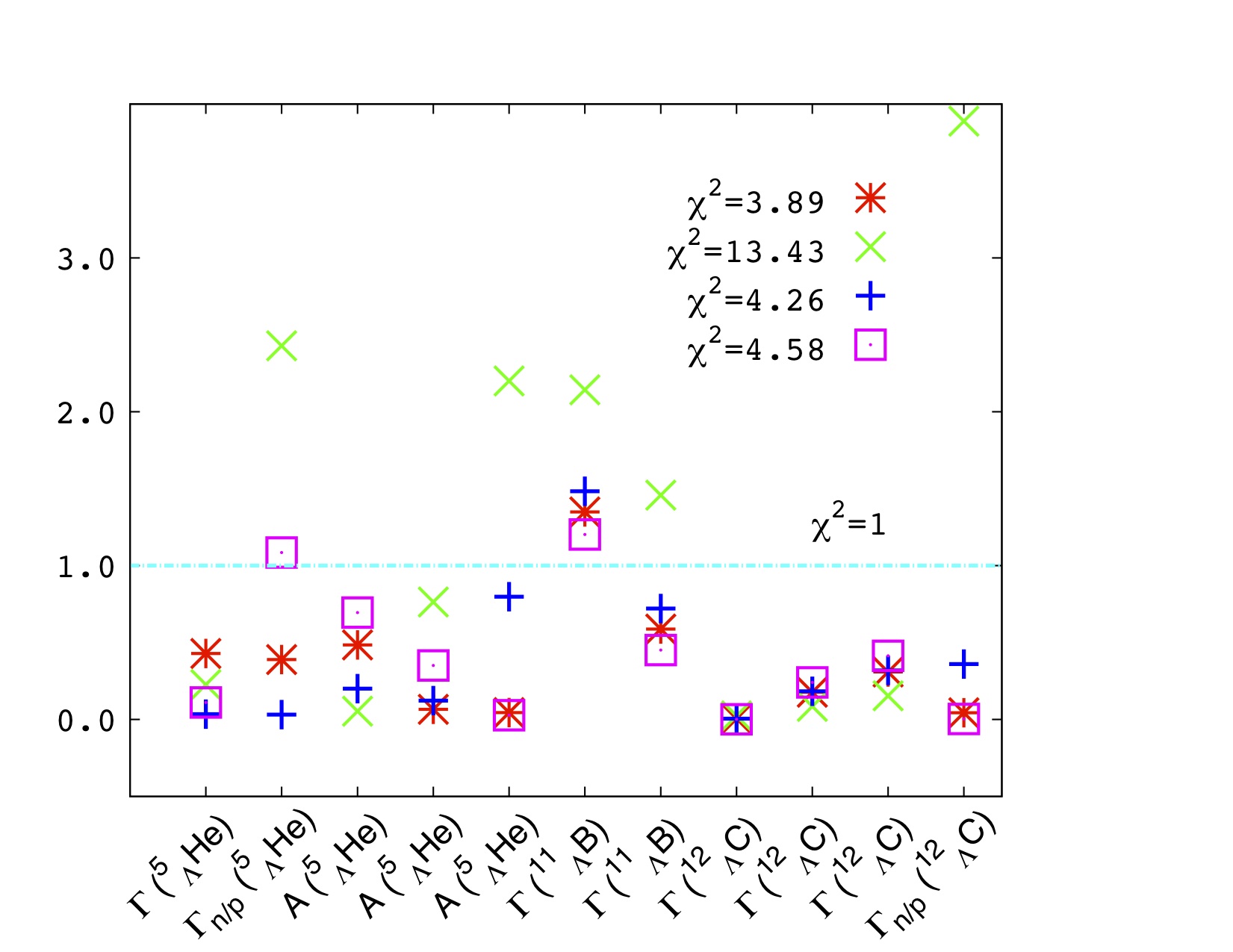}
\caption{Contribution of each experimental point included in the fit to the total $\chi^2$  for the four different fits discussed in the text.}
\label{chips}
\end{center}
\end{figure}

The fits give two minima for each one of the strong interaction
models. Note that the two models differ not only on the kaon exchange
contribution (coupling constants and cut-offs), but also on the
final NN wave functions. The corresponding total $\chi^2$ for a fit to 11 observables is also given in the table. In Fig.~\ref{fits} we show the values for the observables 
used in the present fit together with their respective fitted values, while Fig.~\ref{chips} shows the contribution of each point to the $\chi^2$.
\ignore{and none of them is compatible with the LECs extracted from the OME expansion.}
\ignore{the following interesting features: 
(i) the meson exchange contributions considered here do
not give any strength to the PC central spin-dependent $C_0^1$ and the
parity-violating $C_1^0$ coefficients; (ii) the central spin-independent 
$C_0^0$ LEC predicted by the EFT is larger than the one given by the OME
model; (iii) the $C_1^1$ EFT coefficient is mainly dominated by its
isoscalar component, in agreement with its OME counterpart, but
without falling into the {\it allowed} error-band range; (iv) the
$C_1^2$ EFT LEC is much larger than its OME partner.
To enlighten the situation we
produce two different EFT fits, where we explictly force  $C_0^1=C_1^0=0$ and
$C_1^0=0$ respectively. The result of such fits are given 
in Table~\ref{tab:lecs2}.}

The results in Table~\ref{tab:lecs} show two important features. First, 
the LECs derived from the two OME models considered, J\"ulich and Nijmegen, 
are compatible albeit mostly due to the large theoretical uncertainties. The 
OME prediction for $C_0^1$ is in both cases compatible with zero. Secondly, 
the comparison between the OME extracted LECs values and the LO PC fitted ones 
shows only partial agreement. The largest disagreement is seen in $C_0^0$ 
in all cases. In the next section we will discuss how this disagreement can be reduced with the inclussion of a scalar exchange in the OME formalism. 

Note that the results for the LECs presented here are different from the
ones given in Ref.~\cite{PBH05}. This comparison has to be made with the results obtained with the Nijmegen NSC97f strong interaction model, which
is the only one used in~\cite{PBH05}. Apart from small (kinematical) changes in the final NN wave functions, and in the regularization of the OKE mechanism, the main difference between both 
calculations resides in the experimental values used to perform the fit.
We have updated our data set in order to include the recent rates extracted from the measure in coincidence of the two nucleons in the final state. Moreover, values of the n/p ratio close to one have been disregarded, following the last experimental and theoretical analysis, and more accurate data with smaller error bars have been included.

\ignore{
The good quality of the fits 
due to the scarce number of data points to be fitted its relevance 
must be regarded with caution. The good quality of the two extra fits 
(which have two and one less parameters than the original fit) basicly 
points out that more experimental data are needed to disentangle 
the different contributions entering in Eq.~\ref{eq:vnnlo}. 

Two appreciations are in order which become apparent from the results 
in tables~\ref{tab:lecs} and~\ref{tab:lecs2}. The run when the two 
parameters are constrained, $C_0^1=C_1^0=0$, which is motivated by 
the OME is the one that gives LEC which are in better agreement with 
the ones computed from the OME. In particular when considering the 
$C_1^2$. This LEC which is the one associated with the 
$\sigma_1 \times \sigma_2) \vec{q}$ term has is only 
}
\begin{widetext}
\vspace*{0.5cm}
\begin{table}[h!]
\begin{center}
 \begin{tabular}{l|r|r|r|r|r|r|}
\cline{2-7}
&\multicolumn{3}{c|}{Nijmegen} &  \multicolumn{3}{c|}{J\"ulich} \\
\cline{2-7}
& \mc{1}{|c|}{OME expansion} & \multicolumn{2}{c}{LO PC
    calculation} &
    \mc{1}{|c|}{OME expansion} & \multicolumn{2}{c|}{LO PC calculation} \\
\hline
\multicolumn{1}{|c|}{$C^0_0$} & $1.07\pm0.88$  & $(-0.92 \pm 0.31)$ &
$(4.01\pm 0.23)$  &  $-1.7\pm2.6$ & $( 4.03 \pm 0.50)$ &  $(0.89 \pm 0.58)$  \\
\multicolumn{1}{|c|}{$C^1_0$} & $0.02\pm0.36$  & $( -2.41 \pm 0.11)$  &
$(0.02 \pm 0.33)$ & $0.12\pm0.37$ & $(-0.30 \pm 0.28)$ & $(-1.52 \pm 0.18)$  \\
\hline
\multicolumn{1}{|c|}{${\chi}^2$}  & & $3.89$ & $13.43$ &  & $4.26$ & $4.58$\\
\hline
\end{tabular}
\end{center}
\caption{Values for the LECs obtained from the two sources: OME
expansion and LO (PC) EFT calculation, using the Nijmegen and J\"ulich
strong interaction models. All the quantities are in units of $G_F = 1.166 \times 10^{-11}$ MeV$^{-2}$.
\label{tab:lecs}}
\end{table}
\end{widetext}

\section{Scalar exchange interaction}

By inspecting Table~\ref{tab:lecs} one clearly sees that the largest discrepancy
affects the $C_0^0$ coefficient.
This could be an indication of the
relevance of a scalar exchange ($\sigma$) which is not explicitly
included in the meson exchange formalism employed. 
A sensible way of inferring qualitatively the physical properties of such scalar would be to add it to the meson exchange description.
The one scalar-exchange (OSE) contribution can be obtained from
the following weak and strong vertices~\cite{SIO05}:
\begin{eqnarray}
 {\cal H}^{\rm S }_{\rm {\sst NN} \sigma}&=&
 \im \, g_{\rm {\sst NN} \sigma} \, \overline {\psi}_{\rm N}
\phi^\sigma \psi_{\rm N} \ , \, \, \nonumber \\
{\cal H}^{\rm W }_{\rm {\sst \Lambda N} \sigma}&=&
\im \, G_F m_\pi^2 \, \overline{\psi}_{\rm N} \,\,
\left( A_\sigma + B_\sigma \gamma_5 \right )
\phi^\sigma \psi_\Lambda \, \left( ^0_1 \right)  \, ,
\end{eqnarray}
where $A_\sigma$ and $B_\sigma$ parametrize the parity conserving
and parity violating weak amplitudes. In the
non-relativistic approximation, the corresponding potential reads,
\begin{equation}
{V_{\rm OSE}} ({\vec q}\,) = - G_F m_\pi^2 \, g_{\sst \rm{NN} \sigma}
\left(
A_\sigma
+
{B_\sigma \over 2 M_W} \vec{\sigma}_1 \vec{q} \right)
.
\label{eq:sigma}
\end{equation}

We can now try to establish the values of the weak couplings $A_{\sigma}$ and $B_{\sigma}$ by direct comparison to the results of the fits. We can obtain information about $A_\sigma$ using the numbers obtained in our LO (parity conserving) fit. Insight on $B_\sigma$ would require a NLO fit, which, as we already mentioned, is not needed to get a reasonable fit to our observables.

The OSE gives contribution, in particular,  to $C_{0}^0$\ignore{ and
  to the $C_1^0$}, which now
becomes:
\begin{eqnarray}
C_0^{0~(\sigma)}&=&
\label{c00s}
     \left[ \,
     \frac{ \vcks}{{m_{\sst K^*}}^2 }\left(\frac{\ckspcv}{2}+\dkspcv\right)
    \, + \,
     \frac{ \vco \, \alfao}{{m_{\omega}}^2 } 
\right .\\\nonumber
&-&\left .
       \frac{\vcks\,\ckspcv}{2{m_{\sst K^*}}^2}
      \, - \,
       \frac{ \vcr \, \alfar}{{m_{\rho}}^2 }
- \frac{A_\sigma g_{\sst \rm{NN}\sigma}}{m_\sigma^2}
\right]\, {m_{\pi}}^2 \,\, {\rm .}
\end{eqnarray}

\ignore{
\begin{eqnarray}
C_{0~sc}^0&=&
\label{c00s}
     \left[ \,
     - \frac{A_\sigma g_{\sst \rm{NN}\sigma}}{m_\sigma^2} \,+\,
     \frac{ \vcks}{{m_{\sst K^*}}^2
     }\left(\frac{\ckspcv}{2}+\dkspcv\right)\right .
    \nonumber\\ &+&
    \left .
     \frac{ \vco \, \alfao}{{m_{\omega}}^2 }
            \, \right] \, {m_{\pi}}^2\,\rm{.}
\end{eqnarray}
}

\ignore{\begin{eqnarray}
C^0_0 &=&
     \left( \,
     - \frac{A_\sigma g_{\sst \rm{NN}\sigma}}{m_\sigma^2}
     + \frac{ \vcks \, \alfaks }{{m_{\sst K^*}}^2 }
    \, + \,
     \frac{ \vcr \, \alfar }{{m_{\rho}}^2 }
    \, + \, \right.
\label{c00s}
\nonumber \\
   & &  \left. \frac{ \vco \, \alfao }{{m_{\omega}}^2 }
     \, \right) \, {m_{\pi}}^2
\nonumber \\
C_1^0&=& -B_\sigma g_{\rm{NN}\sigma} {m_\pi^2 \over m_\sigma^2} \,
\rm{.}
\end{eqnarray}}

Since $C_{0}^1$ is not modified by the inclusion of the $\sigma$, the
minima that may be improved via this mechanism are the ones in which
this coefficient is already in agreement with the one obtained from the OME expansion. Focusing
on these minima (the ones with $\chi^2=13.43$ and $\chi^2=4.26$), we can
extract the value of $A_\sigma$ needed to make \ignore{them compatible with
the OME LECs.} the two formalisms
agree (at LO) within each strong interaction model. Using $m_\sigma=550$ MeV and $g_{NN\sigma}=8.8$ \cite{Ma89}
 we get values for $A_\sigma$ in the range $3.3 \to 7.3$ for the Nijmegen minimum and in the range $4.8 \to 16$ for
the J\"ulich one. 

\begin{figure}[h!]
\begin{center}
\hspace*{-0.3cm}
\includegraphics[scale=0.33]{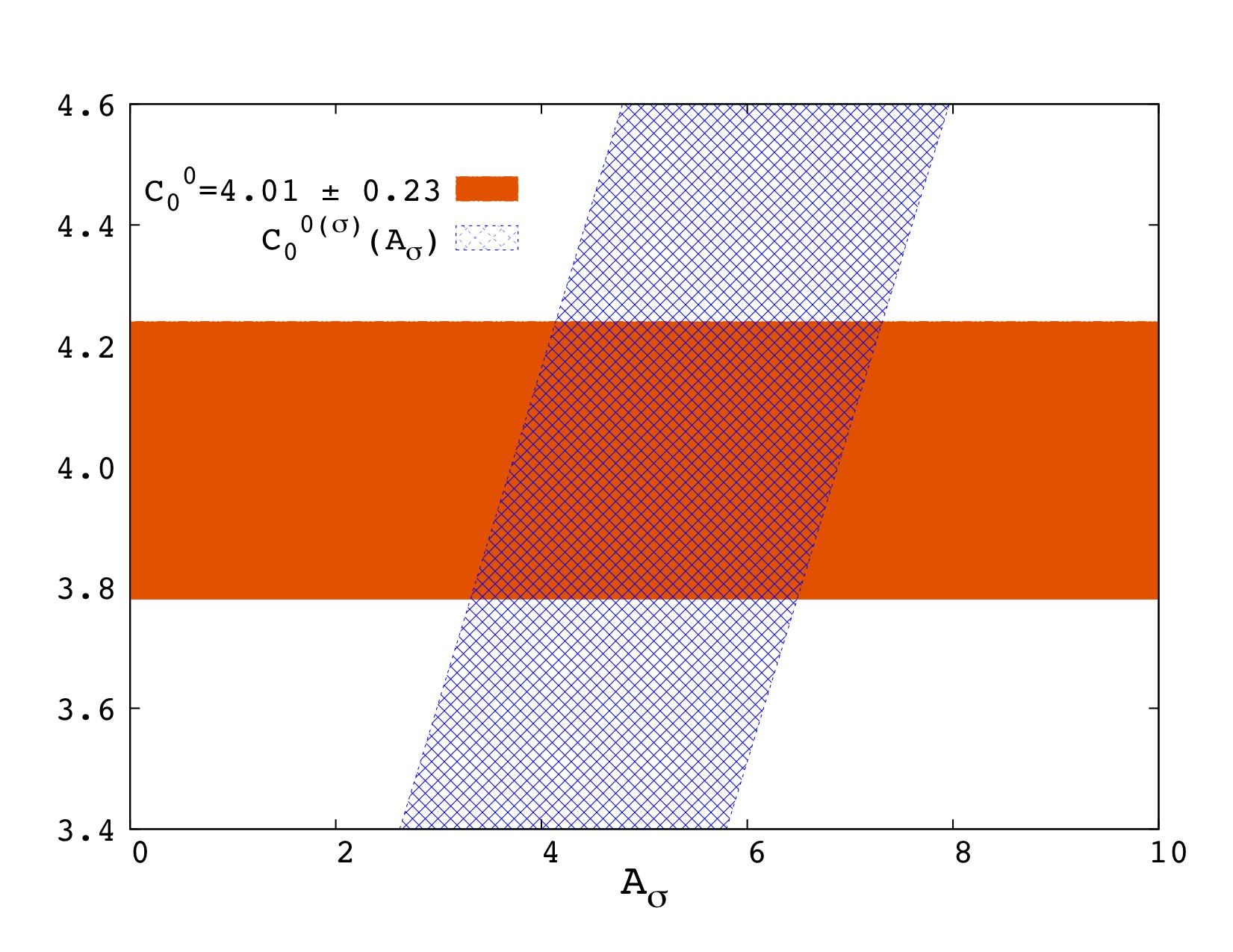}
\caption{Comparison between $C_0^0$ and $C_0^{0~(\sigma)}$ for the
  Nijmegen minimum. The shaded blue area represents the dependence of $C_0^{0~(\sigma)}$ on $A_\sigma$ given by Eq.~(\ref{c00s}), while the 
fitted EFT $C_0^0$ value is respresented by the solid orange area. See text for details.}
\label{fig:sigman}
\end{center}
\end{figure}

\begin{figure}[h!]
\begin{center}
\hspace*{-0.3cm}
\includegraphics[scale=0.33]{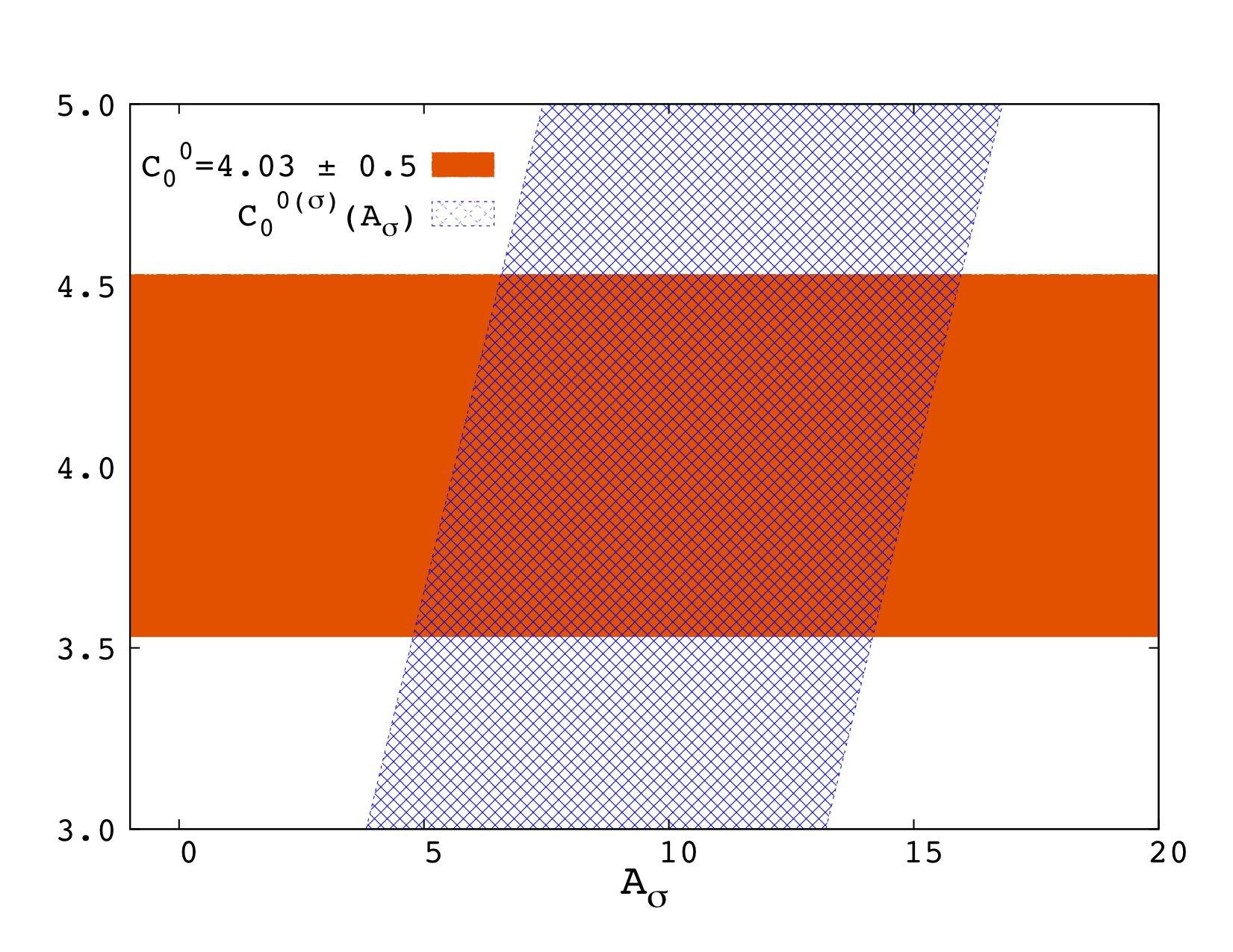}
\caption{Same as Fig.~\ref{fig:sigman} but for the 
  J\"ulich minimum.}
\label{fig:sigmaj}
\end{center}
\end{figure}

The shaded blue band in Fig.~\ref{fig:sigman} (\ref{fig:sigmaj}) shows the value of $C_0^{0~(\sigma)}$ given by Eq.~(\ref{c00s}) as a
function of $A_\sigma$, when the Nijmegen (J\"ulich) strong interaction
model is used. Note that the error band in $C_0^{0~(\sigma)}$ is given by the propagation of the uncertainties in the baryon-baryon-meson coupling constants, taken to be of the order of $30\%$. In the same plot we represent the corresponding fitted value in the EFT approach (solid orange band). The range for $A_\sigma$ quoted before corresponds to the intersection of both bands in the plot, {\it i.e}, the
values for $A_\sigma$ that make compatible 
the OME and EFT formalisms.

Other works have fitted this same
parameter using different approaches.
For instance, Ref.~\cite{SIO05}, which incorporates the OPE, OKE and OSE mechanisms together with a direct-quark transition, uses the phenomenological approach of Block and Dalitz~\cite{BD63} to write the nonmesonic decay rates in terms of the squares of the amplitudes given in Table~\ref{tab:pw} for the s-shell $^5_\Lambda$He, 
 $^4_\Lambda$He and $^4_\Lambda$H hypernuclei. This factorization in terms of two-body amplitudes is possible when effective (spin-independent) correlations are used to account for the strong interaction 
among baryons, where no mixing between the different partial waves is possible. The strong interaction model used in this work is NSC97f. This approach leads to a quadratic equation to determine the couplings, resulting in two values for $A_\sigma$, $3.9$ and $-1.0$ (note that the first of these two values is compatible with the range we are quoting for this constant when the same strong interactions model is used).
Another approach was followed in Ref.~\cite{BM06}, where the exchanges of all the mesons belonging to the pseudoscalar and vector mesons octets are considered in the weak transition in addition to the $\sigma$ meson, while again, effective (spin independent) correlations are used in the strong sector. 
Fixing the value of the strong $NN\sigma$ coupling to be the same as the 
$\pi NN$ one, a range of variation for the $\sigma$ mass and cut-off leads to different values for the weak couplings, once a fit to the nonmesonic decay rate and the neutron-to-proton ratio for $^5_\Lambda$He is performed. Even though the inclusion of the $\sigma$ exchange mechanism does modify their prediction for the intrinsic asymmetry, their results are insensitive to the particular values of the $A_\sigma$ and $B_\sigma$ couplings, and a simultaneous reproduction of all the data is not achieved. 

\section{Conclusions and summary}

We have derived the relations between the low energy coefficients appearing in the EFT description of the two-body $\Lambda N \to NN$ transition driving the decay of hypernuclei and the parameters appearing in
the widely used meson-exchange model. This has been achieved by comparing
the momentum space expansion of the OME potentials to the different orders in the EFT formalism.

In both approaches, the one-pion- and one-kaon-exchange mechanisms are
explicitly included to account for the long and intermediate ranges
of the interaction. The higher mass contributions ($\eta$, $\rho$, $\omega$
and $K^*$) in the OME model are parametrized as contact four-point 
interactions in the EFT approach. 
With this procedure we obtain relations for
the LECs in terms of the masses, couplings and cut-offs characteristic of 
the OME formalism. The numerical values for the LO EFT LECs have been obtained by fitting the available experimental
data for hypernuclear decay observables. In the OME case,
however, the LECs have been written in terms of the masses,
couplings and cut-offs, taken from their experimental values,
symmetry constraints or strong interaction models.

The considered experimental database of hypernuclear decay 
observables can be described with good accuracy within a 
LO EFT supplemented by $\pi$ and $K$ meson exchanges. This 
implies that further experimental efforts will be needed to 
constrain the higher order terms in the EFT of hypernuclear decay. 

Finally, we have analyzed the contribution of a scalar exchange in OME models,
estimating the size of the corresponding parity conserving amplitude, needed
to achieve a better agreement to the available experimental data.

\ignore{giving constraints on both the parity
violating and parity conserving pieces of such scalar exchange.}

\begin{acknowledgments}
We thank Angels Ramos and Joan Soto for a careful reading of the
manuscript.

This work was partly supported by the contract FIS2008-01661 from
MEC (Spain) and FEDER, the EU contract FLAVIAnet MRTN-CT-2006-035482, by the 
contract MIUR 2001024324\_007, by the contract 2005SGR-00343 from 
the Generalitat de Catalunya and by the European Community
Research Infrastructure Integrating Activity ``Study
of Strongly Interacting Matter" (acronym Hadron-
Physics2, Grant Agreement n. 227431) under the
7th Framework Programme of the EU. 
A. P\'erez-Obiol acknowledges the APIF PhD scholarship program of the University of Barcelona.

\end{acknowledgments}
\appendix
\section{Meson-exchange potentials}
\label{appA}

The weak and strong vertices entering the one-pion-exchange (OPE)
amplitude are:
\begin{eqnarray}
{\cal H}^{\rm W }_{\rm {\Lambda N} \pi}&=& \im G_F m_\pi^2
\overline{\psi}_{\rm N}
(A_\pi+B_\pi \gamma_5)
{\vec \tau} \, {\vec \phi}^{\, \pi}
\psi_\Lambda \, \left( ^0_1 \right) \ ,
\nonumber \\
{\cal H}^{\rm S }_{\rm {NN} \pi}&=& \im \, g_{\rm
 {NN} \pi} \,
\overline{\psi}_{\rm N}
\gamma_5
{\vec \tau} \, {\vec \phi}^{ \, \pi} \psi_{\rm N} \ ,
\end{eqnarray}
where $G_F m_\pi^2= 2.21\times 10^{-7}$ is the weak coupling
constant, and $A_\pi$ and $B_\pi$, empirical constants adjusted
to the observables of the free $\Lambda$ decay, which determine
the strength of the parity-violating and parity-conserving
amplitudes, respectively. The nucleon, lambda and pion fields
are given by $\psi_{\rm N}$, $\psi_\Lambda$ and ${\vec \phi}^{\, \pi}$,
respectively, while the isospin spurion $\left( ^0_1 \right)$ is included
to enforce the empirical $\Delta T=1/2$ rule observed in the decay
of a free $\Lambda$. The Bjorken and Drell convention for the
definition of $\gamma_5$~\cite{BD64} is taken.

For the exchange of the pseudoscalar $\eta$ and $K$ mesons, the
strong and weak vertices are (weak constants
are given in units of $G_F m_\pi^2$) :
\begin{eqnarray}
{\cal H}^{\rm S }_{\rm {\sst NN} \eta}&=&
 \im \, g_{\rm {\sst NN} \eta} \, \overline {\psi}_{\rm N}
 \gamma_5
\phi^\eta \psi_{\rm N} \ , \, \, \nonumber \\
{\cal H}^{\rm W }_{\rm {\sst \Lambda N} \eta}&=&
\im \, \overline{\psi}_{\rm N} \,\,
(A_\eta+B_\eta \gamma_5)
\phi^\eta \psi_\Lambda \, \left( ^0_1 \right)  \ , \, \, \nonumber  \\
{\cal H}^{\rm S }_{\sst \Lambda {\rm N} K} &=&
\im \, g_{\sst \Lambda {\rm N} K} \, \overline{\psi}_{\rm N}
\gamma_5 \,\,
\phi^{\sst \rm K} \psi_\Lambda  \ , \, \,
\nonumber \\
{\cal H}^{\rm W }_{\sst {\rm NN} K} &=&
\im \, \left[ \, \overline{\psi}_{\rm N} \left( ^0_1 \right)
\,\,( C_{\rm \sst{K}}^{\rm \sst{PV}} +
C_{\rm  \sst{K}}^{\rm \sst{PC}}
\gamma_5) \,\,(\phi^{\sst \rm K})^\dagger
\psi_{\rm N} \right.
 \\ \nonumber
& & \left. + \, \overline{\psi}_{\rm N} \psi_{\rm N}
\,\,( D_{\rm \sst{K}}^{\rm\sst{PV}}
+ D_{\rm\sst{K}}^{\rm\sst{PC}}
\gamma_5 ) \,\,(\phi^{\sst \rm K})^\dagger \,\,
\left( ^0_1 \right) \right] \ ,
\label{eq:kaweak}
\end{eqnarray}
where the weak coupling constants cannot be taken directly from
experiment.

The weak $\Lambda {\rm N} \rho$, $\Lambda \rm N \omega$, ${\rm NN} K^*$ and
strong NN$\rho$, NN$\omega$, $\Lambda {\rm N} K^*$ vertices are given
by~\cite{MG84}:

\begin{eqnarray}
{\cal H}^{\rm W}_{\rm {\sst \Lambda N} \rho} &=&
{\overline \psi}_{\rm N} \: \bigg( \alpha_\rho \gamma^\mu   \\
&& \phantom{ \bigg(}  - \beta_\rho \im \frac{\sigma^{\mu \nu} q_\nu} {2 \overline{M}} +
\varepsilon_\rho
\gamma^\mu \gamma_5 \bigg)
{\vec \tau} \, {\vec \rho}_\mu \,\psi_\Lambda \,\,
\left( ^0_1 \right)  \ , \, \, \nonumber \\
{\cal H}^{\rm S}_{\rm {\sst NN} \rho} &=& \overline{\psi}_{\rm N}
\left( g^{\rm {\sst V}}_{\rm {\sst NN} \rho}
 \gamma^\mu + \im
\frac{ g^{\rm {\sst T}}_{\rm {\sst NN} \rho}}{2M} \sigma^{\mu \nu}
q_\nu \right) {\vec \tau} \, {\vec \rho}_\mu \, \psi_{\rm N} \ , \, \,
\label{eq:rhohamil}
\end{eqnarray}

\begin{eqnarray}
{\cal H}^{\rm S }_{\rm {\sst NN} \omega}&=&
\overline{\psi}_{\rm N} \left( g^{\rm {\sst V}}_{\rm
 {\sst NN} \omega}
\gamma^\mu + \im \frac{g^{\rm {\sst T}}_{\rm {\sst NN}
 \omega}}{2M}
\sigma^{\mu \nu} q_\nu \right)
\phi^\omega_\mu \psi_{\rm N}  \\
{\cal H}^{\rm W}_{\rm {\sst \Lambda N} \omega} &=&  \:
{\overline \psi}_{\rm N}
\:
\bigg( \alpha_\omega \gamma^\mu  \\
&& \phantom{\bigg(}
- \beta_\omega \im \frac{\sigma^{\mu \nu} q_\nu} {2 \overline M} +
\varepsilon_\omega
\gamma^\mu \gamma_5 \bigg)
\phi^\omega_\mu \psi_\Lambda \left( ^0_1 \right)  \ , \, \, \nonumber \\
{\cal H}^{\rm S }_{\rm {\sst \Lambda N K^*}}&=&
\overline{\psi}_{\rm N} \left(g^{\rm \sst{V}}_{\rm
 \sst{\Lambda N K^*}}
\gamma^\mu + \im \frac{g^{\rm \sst{T}}_{\rm
 \sst{\Lambda N K^*}}}{2
\overline M}
\sigma^{\mu \nu} q_\nu \right)
\phi^{\rm\sst{K^*}}_\mu \psi_\Lambda , \, \,
\\
{\cal H}^{\rm W}_{\rm\sst{NN K^*}} &=& \: \left( \,\,
C_{\rm\sst{K^*}}^{\rm \sst{PC,V}} \overline{\psi}_{\rm N} \left(^0_1 \right)
(\phi^{\rm\sst{K^*}}_\mu)^\dagger\,\,
\gamma^\mu \psi_{\rm N}  \right.
\nonumber \\
&+&
D_{\rm\sst{K^*}}^{\rm\sst{PC,V}}
 \overline{\psi}_{\rm N} \gamma^\mu
\psi_{\rm N}
(\phi^{\rm\sst{K^*}}_\mu)^\dagger \,\,
\left( ^0_1 \right)
\nonumber \\
&+&
 C_{\rm\sst{K^*}}^{\rm\sst{PC,T}}
\overline{\psi}_{\rm N}
\left( ^0_1 \right) (\phi^{\rm\sst{K^*}}_\mu)^\dagger\,\,
(- \im) \frac{\sigma^{\mu \nu} q_\nu} {2 M}
\psi_{\rm N}
\nonumber \\
&+&
D_{\rm\sst{K^*}}^{\rm\sst{PC,T}}
 \overline{\psi}_{\rm N}
(- \im) \frac{\sigma^{\mu \nu} q_\nu} {2 M}
\psi_{\rm N}
(\phi^{\rm\sst{K^*}}_\mu)^\dagger \,\,
\left( ^0_1 \right)  \nonumber \\
&+&
C_{\rm\sst{K^*}}^{\rm\sst{PV}}
 \overline{\psi}_{\rm N} \left(
^0_1 \right)
(\phi^{\rm\sst{K^*}}_\mu)^\dagger\,\,
\gamma^\mu \gamma_5 \psi_{\rm N}
\nonumber \\
&+& \left.
D_{\rm\sst{K^*}}^{\rm\sst{PV}} \overline{\psi}_{\rm
 N} \gamma^\mu
\gamma_5 \psi_{\rm N}
(\phi^{\rm\sst{K^*}}_\mu)^\dagger \,\,
\left( ^0_1 \right)  \,\, \right) \ .
\end{eqnarray}

\section{LECs in terms of meson-exchange parameters}
\label{appB}

The expressions for the LECs in terms of the meson
exchange parameters are the following:
\begin{eqnarray}
C_{0~sc~}^0&=&
     \left[ \,
     \frac{ \vcks}{{m_{\sst K^*}}^2 }\left(\frac{\ckspcv}{2}+\dkspcv\right)
    \, + \,
     \frac{ \vco \, \alfao}{{m_{\omega}}^2 } \right]\, {m_{\pi}}^2
      \ , \, \, \nonumber \\ \nonumber\\
C_{0~vec}^0&=& \, \left(
       \frac{\vcks\,\ckspcv}{2{m_{\sst K^*}}^2}
      \, + \,
       \frac{ \vcr \, \alfar}{{m_{\rho}}^2 }
       \, \right) \, {m_{\pi}}^2
       \ , \, \, \nonumber \\ 
C_{0~sc~}^1&=&0
       \ , \, \, \nonumber \\ 
C_{0~vec}^1&=&0
        \ , \, \, \nonumber \\
C^0_{1~sc~} &=&  0
      \ , \, \, \nonumber \\
C^0_{1~vec} &=&  0
      \ , \, \, \nonumber \\
C^1_{1~sc~} &=&
\frac{-{m_{\pi}}^2}{2 M} 
      \frac{\, A_\eta \, g_{\sst {\rm NN} \eta}}
{{m_\eta}^2 } \,
      \ , \, \, \nonumber \\
C^1_{1~vec} &=&0  \ , \, \,
\end{eqnarray}

\begin{widetext}
\begin{eqnarray}
C_{1~sc~}^2&=&
     \,-\,
     \frac{{m_{\pi}}^2}{2 M}
      \left[
      \frac{{\rm i} \, ( \vcks + \tcks ) \,(\frac{\ckspv}{2}+\dkspv) \, {m_{\pi}}^2 \, }
           { {m_{\sst K^*}}^2 }  \, + \,
      \frac{{\rm i} ( \vco + \tco ) \epso \, {m_{\pi}}^2 \, }
       { {m_{\omega}}^2 }
       \right]
       \ , \, \, \nonumber\\
C_{1~vec}^2&=&
     \frac{{m_{\pi}}^2}{2 M}
      \left[
      \frac{{\rm i} \, ( \vcks + \tcks ) \,\ckspv \, {m_{\pi}}^2 \, }
           { {2m_{\sst K^*}}^2 }  \, + \,
      \frac{{\rm i} \, ( \vcr + \tcr ) \epsr \, {m_{\pi}}^2 \, }
       { {m_{\rho}}^2 }
      \right]
       \ , \, \, \nonumber\\\nonumber
C_{2~sc~}^0&=&
         \frac{{m_\pi}^2}{4M\overline{M}}
         \left[\left(
         \frac{\ckspcv}{2}+\dkspcv+\frac{\ckspct}{2}+\dkspct\right)\frac{\vcks+\tcks}{{m_{\sst K*}}^2}
         \,+\,
         \frac{\left(\alfao+\betao\right)\left(\vco+\tco\right)}{{m_\omega}^2}-\frac{B_\eta\,\ce}{{m_\eta}^2}
          \right]  \ , \, \,
          \\\nonumber 
C_{2~vec}^0&=&
        \frac{{m_\pi}^2}{4M\overline{M}}
        \left[\frac{\left(\ckspcv+\ckspct\right)\left(\vcks+\tcks\right)}{2{m_{\sst K*}}^2}
         \,+\,
         \frac{\left(\alfar+\betar\right)\left(\vcr+\tcr\right)}{{m_\rho}^2}\right]  \ , \, \,
          \\\nonumber 
C_{2~sc~}^1&=&
         -\frac{{m_\pi}^2}{4M\overline{M}}
         \left[\left(
         \frac{\ckspcv}{2}+\dkspcv+\frac{\ckspct}{2}+\dkspct\right)\frac{\vcks+\tcks}{{m_{\sst K*}}^2}
         \,+\,
         \frac{\left(\alfao+\betao\right)\left(\vco+\tco\right)}{{m_\omega}^2}
          \right]  \ , \, \,
          \\\nonumber 
C_{2~vec}^1&=&
        -\frac{{m_\pi}^2}{4M\overline{M}}
        \left[\frac{\left(\ckspcv+\ckspct\right)\left(\vcks+\tcks\right)}{2{m_{\sst K*}}^2}
         \,+\,
         \frac{\left(\alfar+\betar\right)\left(\vcr+\tcr\right)}{{m_\rho}^2}\right]  \ , \, \,
          \\
C_{2~sc~}^2&=&
         \,-\,2{m_\pi}^2\left[\frac{\vcks\,\left(\Lambda^2+{m_{\sst
                 K^*}}^2\right)
             \left(\frac{\ckspcv}{2}+\dkspcv\right)}{{m_{\sst
                 K^*}}^4\Lambda^2}
           \,+\,\frac{\vco\alfao\left(\Lambda^2+{m_\omega}^2\right)}{{m_\omega}^4\Lambda^4}\right]  \ , \, \,
          \\\nonumber 
C_{2~vec}^2&=&
          \,-\,2{m_\pi}^2
           \left[\frac{\vcks\,\left(\Lambda^2+{m_{\sst
                 K^*}}^2\right)
             \,\ckspcv}{{2m_{\sst
                 K^*}}^4\Lambda^2}
           \,+\,\frac{\vcr\alfar\left(\Lambda^2+{m_\rho}^2\right)}{{m_\rho}^4\Lambda^4}\right]  \ . \, \,
\end{eqnarray}

\end{widetext}

\end{document}